\documentclass[prd,12pt]{revtex4}

\usepackage{epsfig}
\usepackage{graphicx}
\usepackage{dcolumn}
\usepackage{amsmath}
\usepackage{latexsym}
\usepackage{color}
\usepackage{subfig}

\usepackage{footnote}
\usepackage{hyperref}
\hypersetup{
    colorlinks=true,
    linkcolor=blue,
    citecolor=blue,
    filecolor=blue,      
    urlcolor=blue,
    hyperfootnotes=true
}

\begin{document}

\title{$p$-wave holographic superconductors with massive vector condensate in Born-Infeld electrodynamics}  

\author{Ankur Srivastav\footnote{
\href{mailto:ankursrivastav@bose.res.in}{ankursrivastav@bose.res.in}},
Debabrata Ghorai\footnote{
\href{mailto:debabrataghorai@bose.res.in}{debabrataghorai@bose.res.in}},
 Sunandan Gangopadhyay\footnote{\href{mailto:sunandan.gangopadhyay@gmail.com}{sunandan.gangopadhyay@gmail.com}\\       \href{mailto:sunandan.gangopadhyay@bose.res.in}{sunandan.gangopadhyay@bose.res.in}}}
\affiliation{Department of Theoretical Sciences, S. N. Bose National Centre for Basic Sciences,\\ Block-JD, Sector-III, Salt Lake City,\\ Kolkata 700106, India}

\begin{abstract}

\noindent In this paper, we have studied the effect of Born-Infeld electrodynamics in holographic $p$-wave superconductors with massive vector condensation. We have analysed this model in the probe limit using a variational method known as the St\"urm-Liouville eigenvalue approach. For this $p$-wave holographic superconductor model, we have calculated the critical temperature $T_{c}$ as well as the value of the condensation operator for two different choices of $m^{2}$. We have also pointed out the similarties and dissimilarities between this model for $m^{2} = 0$ and $p$-wave holographic superconductor model constructed out of Einstein-Yang-Mills theory. We have then computed the conductivity of these holographic superconductor models using a self-consistent approach and have shown that the DC conductivity diverges.

\end{abstract}

\maketitle


\section{Introduction}
\noindent In the last two decades gauge/gravity duality has emerged as a powerful tool to study various condensed matter systems which are strongly correlated \cite{jm, jmetal}. This apparent connection between a gravity theory and a gauge theory was expected for many years in the form of holographic principle and has been precisely conjectured for a particular gauge theory relating classical gravity theory in anti de-Sitter (AdS) spacetime \cite{jm}. Although the conjecture was about a duality between a gravity theory in AdS spacetime and a conformal field theory in one lower dimension spacetime, that is,  at the boundary of AdS spacetime where gravity theory lives, later researches conceded to a more general form of the strong/weak duality between asymptotically AdS spacetime and nearly conformal field theory at the boundary of AdS. Later on, this duality has been utilised to study various physical systems from both sides. However, it turns out that there are many strongly correlated systems in condensed matter physics which are difficult to deal with traditional field theoretic methods. Fortunately, gauge/gravity duality provides us with an opportunity to study such difficult systems via their gravity dual models in one higher dimensional spacetime. These gravitational duals are far easy to deal with as we can study them in the classical general relativistic senario.\\
\noindent Inspired by the simple model of Abelian symmetry breaking around a charged black hole in AdS spacetime proposed in \cite{ssg}, a gravity dual model that mimicked the properties of a s-wave superconductor was developed in \cite{hhh}. Since then so many investigations have been around investigating such gravity duals mimicking various types of superconductors in numerous physical situations \cite{ssp, gks, hr, pwpop, lcz, st, sgdr1, sg1, RGC, rc2, assg}. One particular interesting study in this regard has been to see the effect of nonlinear electrodynamics in these gravity duals. There are many ways to incorporate such nonlinearity in these models but the inclusion of the Born-Infeld (BI) electrodynamics \cite{dbi1, dbi2, dbi3, kru} is of profound interest as it is the only nonlinear theory that has duality symmetry just like ordinary Maxwell electrodynamics. Several studies have been carried out incorporating the effect of BI electrodynamics in holographic superconductors \cite{jc, sgdr, sgdr2, rbsg, sg2, pcgs, dgsg1, dgsg2, dgsg3}. Another motivation to consider the BI electrodynamics comes straight from the string theory where the BI electrodynamics describes the low energy behaviour of D branes \cite{rbsg}.\\
\noindent There is another important gravity model with a charged vector field in the bulk as the vector order parameter that corresponds to the holographic $p$-wave superconductor. Such a model for holographic $p$-wave superconductor using $SU(2)$ Yang-Mills field in the bulk was first provided in \cite{ssp}. In this model a gauge boson generated by one $SU(2)$ generator works as a dual to the vector order parameter. Unlike in a $s$-wave holographic superconductor, here the onset of the condensate spontaneously breaks not only the $U(1)$ symmetry but also $SO(2)$ rotational symmetry in the $x$-$y$ plane \cite{RGC}.\\
\noindent Recently a new gravity dual model has been proposed for the $p$-wave superconductor using a complex vector field non-minimally coupled to the Maxwell field \cite{rc2}. A detailed analysis of the phase diagram for this model was also provided in \cite{rc2}. A similar phase diagram analysis has been done for a slightly modified version of this model where the effect of non-linearity was incorporated in the Maxwell field via Born-Infeld parameter \cite{pcgs}. However, explicit analytic calculations for the condensation and conductivity in this model has not been carried out in the literature. In this paper we have analytically obtained the critical temperature, the condensation operator value and the conductivity for the holographic $p$-wave superconductor model proposed in \cite{rc2} in the presence of Born-Infeld electrodynamics. \\
\noindent We have organised this paper in the following manner. In section II, we have developed the model and have found the equations of motion for the matter field and the gauge field with appropriate ansatz. Then, in section section III, we have used the St{\"u}rm-Liouville method to find the critical temperature and the condensation operator. We have calculated the conductivity for this model in section IV. Finally, we have summarised our findings and draw relevant conclusions in section V. We have performed all our computations in the probe limit, where we can ignore the backreaction of the matter field in the metric.
\section{Set Up for $p$-wave Holographic Superconductors}
\noindent \hypertarget{sec2}{Holographic} superconductors with $p$-wave gap are based on the solutions to field equations of Einstein-Yang-Mills theory with a cosmological constant. The action for this model reads 
\begin{eqnarray}
 \mathcal{S} = \dfrac{1}{2\mathcal{G}^2} \int d^{4}x \bigg(\mathcal{R}-\dfrac{1}{4}(F^{a}_{\mu\nu})^{2}+\dfrac{6}{L^{2}}\bigg)
\label{EYM action}
\end{eqnarray}
where $F^{a}_{\mu\nu}$ is the field strength tensor of an $SU(2)$ gauge field.\\
\noindent \hypertarget{sec2}{We} work with the metric of a planar black hole in $AdS_{3+1}$ spacetime arising from the solution of Einstein gravity 
\begin{eqnarray}
 ds^{2}=-f(r) dt^{2}+r^{2}(dx^{2}+dy^{2})+\dfrac{dr^{2}}{f(r)}
\label{metric}
\end{eqnarray}
where $$f(r)=\bigg(r^{2}-\dfrac{r_0^{3}}{r}\bigg) $$ with $r_{0}$  being the event horizon of the black hole, and the AdS radius has been set to unity. The Hawking temperature associated with the above black hole geometry is given by 
\begin{eqnarray}
T=\dfrac{3r_{0}}{4\pi}~.
\label{Hawking_temperature}
\end{eqnarray}

\noindent We now write down the model for holographic $p$-wave superconductor with the Lagrangian density consisting of a Maxwell field $A_{\mu}$ and a massive complex vector field $\rho_{\mu}$.  The action for this model reads
\begin{eqnarray}
\mathcal{S} = \dfrac{1}{16\pi G}\int d^{4}x \sqrt{-g}\bigg(\mathcal{R}-2\Lambda +\mathcal{L}\bigg)
\label{Action}
\end{eqnarray}
where
\begin{eqnarray}
~~~~~~~~~ \mathcal{L}=\dfrac{1}{b}\bigg(1-\sqrt{1+\dfrac{b}{2}F^{\mu\nu}F_{\mu\nu}}\bigg)-\dfrac{1}{2}\rho_{\mu\nu}^{\dagger}\rho^{\mu\nu}-m^{2}\rho_{\mu}^{\dagger}\rho^{\mu}
\label{Lagrangian_density}
\end{eqnarray}
$$F_{\mu\nu} \equiv \partial_{[\mu}A_{\nu]} ~ , ~ D_{\mu} \equiv (\partial_{\mu}-iA_{\mu}) ~ , ~ \rho_{\mu\nu} \equiv D_{\mu}\rho_{\nu}-D_{\nu}\rho_{\mu}. $$
The Lagrangian density $\mathcal{L}$ consists of Born-Infeld electrodynamics and $b$ is the Born-Infeld parameter. Since the metric $g_{\mu\nu}$ depends only on $r$, we take the following ansatz for the matter field and the gauge field respectively $$\rho_{\mu}=\delta^{x}_{\mu}~\rho(r)~~,~~A_{\mu}=\delta^{t}_{\mu}\Phi(r)~.$$
Now varying the action $\mathcal{S}$ in eq.(\ref{Action}), we get the equations of motion for the matter field $\rho(r)$ and the gauge field $\Phi(r)$
\begin{eqnarray}
   \rho^{\prime\prime}+ \dfrac{f^{\prime}}{f}\rho^{\prime}+\bigg(\dfrac{\Phi^{2}}{f^{2}}-\dfrac{m^{2}}{f}\bigg)\rho=0
\label{eom_rho}
\end{eqnarray}
\begin{eqnarray}
    \Phi^{\prime\prime}+\dfrac{2}{r}\Phi^{\prime}(1-b\Phi^{\prime 2})-\dfrac{2\Phi\rho^{2}}{r^{2}f}(1-b\Phi^{\prime 2})^{3/2}=0 
\label{eom_Phi}
\end{eqnarray}
where prime denotes the derivative with respect to $r$.\\
We now make the change of coordinate,  $z=\dfrac{r_{0}}{r}$, such that the horizon is at $z=1$ while the AdS boundary is at $z=0$. In this coordinate, the  field eq.(s)(\ref{eom_rho}, \ref{eom_Phi}) take the following form
\begin{eqnarray}
\rho^{\prime\prime}-\dfrac{3z^{2}}{(1-z^{3})}\rho^{\prime}+\dfrac{1}{z^{2}(1-z^{3})}\bigg(\dfrac{z^{2}\Phi^{2}}{r_{0}^{2}(1-z^{3})}-m^{2}\bigg)\rho=0
\label{eom_rho_in_z}
\end{eqnarray}
\begin{eqnarray}
\Phi^{\prime\prime}+\dfrac{2bz^{3}}{r_{0}^{2}}\Phi^{\prime3}-\dfrac{2\Phi\rho^{2}}{r_{0}^{2}(1-z^{3})}\bigg(1-\dfrac{bz^{4}}{r_{0}^{2}}\Phi^{\prime 2}\bigg)^{3/2}=0 
\label{eom_Phi_in_z}
\end{eqnarray}
where prime denotes derivative with respect to the new coordinate $z$.\\
From the gauge/gravity duality dictionary, the behaviour of $\Phi(z)$ and $\rho(z)$ near the AdS boundary are known to be of the following form
\begin{eqnarray}
\Phi(z) = \mu - \dfrac{\tilde{\rho}}{r_{0}}z 
\label{phi_asymptote}
\end{eqnarray}
\begin{eqnarray}
\rho(z) \simeq \dfrac{\rho_{+}}{r_{0}^{\Delta_{+}}}z^{\Delta_{+}}+\dfrac{\rho_{-}}{r_{0}^{\Delta_{-}}}z^{\Delta_{-}} 
\label{rho_asymp}
\end{eqnarray}
where $\mu$ is the chemical potential and $\tilde{\rho}$ is the charge density. $\Delta_{\pm}$ are roots of the equation
\begin{eqnarray}
\Delta = \dfrac{1}{2}(1 \pm \sqrt{1+4m^{2}})~.
\label{con_dim}
\end{eqnarray}
Here $\Delta$ is known as the conformal dimension and it depends on $m^{2}$ through the above relation \footnote{This relation can be obtained using eq.(\ref{rho_asymp}) in eq.(\ref{eom_rho_in_z}).}.
It is apparent from eq.(\ref{rho_asymp}) that $\Delta$ must be real and positive. With this condition on $\Delta$, the choice of $m^{2}$ is also restricted. To fulfil the above mentioned condition for $\Delta$, $m^{2}$ needs to satisfy the following lower bound.
\begin{eqnarray}
m^{2} \geq -\dfrac{1}{4}~. 
\label{B_F_bound}
\end{eqnarray}
Eq.(\ref{B_F_bound}) is famously known as the Breitenlohner-Freedman (BF) bound \cite{BF2}. The BF bound implies that the vector field, even if it has negative mass, is stable in AdS spacetime as long as eq.(\ref{B_F_bound}) is satisfied. \\ 
With this set up in hand, we shall proceed to carry out the St\"urm-Liouville analysis in the next section.
\section{St{\"u}rm-Liouville Analysis}
\subsection{Critical Temperature}
\noindent \hypertarget{sec4}{In} this section, we shall apply the St{\"u}rm-Liouville eigenvalue method to find the critical temperature and the value of the condensation operator. We first recall that the matter field $\rho(z)$ vanishes at the critical temperature $T_{c}$. Hence, at $T = T_{c}$, eq.(\ref{eom_Phi_in_z}) simplifies to the following form  
\begin{eqnarray}
\Phi^{\prime\prime}+\dfrac{2bz^{3}}{r_{0}^{2}}\Phi^{\prime 3}=0~.
\label{eom_Phi_at_Tc}
\end{eqnarray}
The analytic solution of eq.(\ref{eom_Phi_at_Tc}) up to first order in the Born-Infeld parameter $b$ is given by \cite{sgdr} 
\begin{eqnarray}
\Phi(z)= \lambda r_{0}(1-z) \bigg[1-\dfrac{b(\lambda^{2}|_{b=0})}{10}\zeta(z)\bigg]
\label{exact_Phi}
\end{eqnarray}
where $\zeta(z)=(1+z+z^{2}+z^{3}+z^{4})$ and $\lambda=\dfrac{\tilde{\rho}}{r_{0c}^{2}}$, $r_{0c}$ being the horizon radius at the critical temperature. From eq.(\ref{Hawking_temperature}), we find the expression for the critical temperature to be   
\begin{eqnarray}
T_{c} = \dfrac{3}{4\pi}\sqrt{\dfrac{\tilde{\rho}}{\lambda}}~.
\label{Critical_temp_eq}
\end{eqnarray}
Now using eq.(\ref{exact_Phi}) in eq.(\ref{eom_rho_in_z}), we  get the following field equation for $\rho$
\begin{eqnarray}
\rho^{\prime\prime}-\dfrac{3z^{2}}{(1-z^{3})}\rho^{\prime}+\dfrac{\lambda^{2}}{(1+z+z^{2})^{2}}\bigg(1-\dfrac{b(\lambda^{2}|_{b=0})}{5}\zeta(z)\bigg)\rho - \dfrac{m^{2}}{z^{2}(1-z^{3})}\rho=0~.
\label{eom_rho_at_Tc}
\end{eqnarray}
To proceed further, we consider the following non-trivial form of the field $\rho(z)$ 
\begin{eqnarray}
\rho(z)=\dfrac{\langle \mathcal{O}_{\Delta} \rangle}{\sqrt{2}r_{0}^{\Delta}}z^{\Delta}F(z)
\label{rho_sturm}
\end{eqnarray}
with the conditions $F(0)=1$ and $F^{\prime}(0)=0$. These boundary conditions on $F(z)$ are consistent with the behaviour of $\rho(z)$ near the AdS boundary given by eq.(\ref{rho_asymp}).
Substituting the form of the field $\rho(z)$ given in eq.(\ref{rho_sturm}) in eq.(\ref{eom_rho_at_Tc}), we obtain
\begin{eqnarray}
\hspace*{-23mm}(z^{2\Delta}(1-z^{3})F^{\prime})^{\prime}-(3\Delta z^{2\Delta +1}+m^{2}z^{2\Delta -2}-\Delta(\Delta -1)z^{2\Delta -2}(1-z^{3}))F \nonumber\\
+\dfrac{\lambda^{2}z^{2\Delta}(1-z)}{(1+z+z^{2})}\bigg(1-\dfrac{b(\lambda^{2}|_{b=0})}{5}\zeta(z)\bigg)F=0~.\hspace*{45mm}~
\label{field_eq_in_F_z}
\end{eqnarray}
Comparing eq.(\ref{field_eq_in_F_z}) with the standard form of the St{\"u}rm-Liouville eigenvalue equation given by  
\begin{eqnarray}
\dfrac{d(p(z)F^{\prime})}{dz}-q(z)F+\lambda^{2}r(z)F=0
\label{Sturm_form_F_z_eq_delta_1}
\end{eqnarray}
we can identify the form of the functions $p(z)$, $q(z)$ and $r(z)$ to be 
\begin{eqnarray}
 p(z) = z^{2\Delta}(1-z^{3}) \hspace*{30mm}~ \nonumber \\ 
 q(z) = 3\Delta z^{2\Delta +1}+m^{2}z^{2\Delta -2}-\Delta(\Delta -1)z^{2\Delta -2}(1-z^{3}) \label{general functions}\\
 r(z) = \dfrac{z^{2\Delta}(1-z)}{(1+z+z^{2})}\bigg(1-\dfrac{b(\lambda^{2}|_{b=0})}{5}\zeta(z)\bigg).\hspace*{10mm}~ \nonumber  
\end{eqnarray}
We can now find the eigenvalue $\lambda^{2}$ in eq.(\ref{field_eq_in_F_z}) from the following relation
\begin{eqnarray}
\lambda^{2} = \dfrac{\displaystyle\int\limits_{0}^{1}dz \big(p(z)F^{\prime 2} + q(z)F^{2}\big)}{\displaystyle\int\limits_{0}^{1}dzr(z)F^{2}}~. 
\label{eigenvalue_eq}
\end{eqnarray}
To estimate $\lambda^{2}$, we choose a trial function for $F(z)$ as $F_{\alpha}(z) = (1-\alpha z^{2})$. The eigenvalue $\lambda^{2}$ is determined by minimizing eq.(\ref{eigenvalue_eq}) with respect to $\alpha$. The value of $\lambda_{\alpha_{min.}}$ can then be used in eq.(\ref{Critical_temp_eq}) to determine the critical temperature of the $p$-wave holographic superconductor from the equation
\begin{eqnarray}
T_{c} = \dfrac{3}{4\pi}\sqrt{\dfrac{\tilde{\rho}}{\lambda_{\alpha_{min.}}}}~.
\label{Critical Temp Min.}
\end{eqnarray}
To move ahead, we select some particular conformal dimension via eq.(\ref{con_dim}). We would focus on the following two choices of $m^{2}$ and its corresponding conformal dimensions $\Delta=(\Delta_{+},~\Delta_{-})$.
\begin{eqnarray}
m^{2} = 0 ~~ \rightarrow ~~\Delta = (1,~0)
\label{m2=0}
\end{eqnarray}
\begin{eqnarray}
m^{2} = -\dfrac{3}{16} ~~ \rightarrow ~~ \Delta = \bigg(\dfrac{3}{4},~\dfrac{1}{4}\bigg)~.
\label{m2=-3/16}
\end{eqnarray}
We know that near the AdS boundary $\rho(z)$ takes the form given by eq.(\ref{rho_asymp}). In order to have spontaneous symmetry breaking, we set the source term $\rho_{-}=0$ for the above choices. Therefore the boundary behaviour of $\rho(z)$ is now given as
\begin{eqnarray}
\rho(z) \simeq \dfrac{\rho_{+}}{r_{0}^{\Delta_{+}}}z^{\Delta_{+}}~.
\label{rho_asymp_positive}
\end{eqnarray}
As the subscript is no longer needed in the above equation, we would simply drop it from now onwards.
\subsubsection*{\textbf{Case(I): $m^{2} = 0,~ \Delta = 1$}}
\noindent In this case, the functions $p(z)$, $q(z)$ and $r(z)$ are obtained by substituting  $m^{2} = 0$ and  $\Delta = 1$ in eq.(\ref{general functions}) and are given as
\begin{eqnarray}
 p(z) = z^{2}(1-z^{3}) \hspace*{40mm}~ \nonumber \\ 
 q(z) = 3 z^{3} \hspace*{53mm}~ \label{m=0 functions}\\
 r(z) = \dfrac{z^{2}(1-z)}{(1+z+z^{2})}\bigg(1-\dfrac{b(\lambda^{2}|_{b=0})}{5}\zeta(z)\bigg). \nonumber  
\end{eqnarray}
Using eq.(\ref{eigenvalue_eq}) with trial function $F_{\alpha}(z) = (1-\alpha z^{2})$ and eq.(\ref{m=0 functions}), the eigenvalue $\lambda^{2}$ reads 
\begin{eqnarray}
\lambda_{\alpha}^{2} = \dfrac{\displaystyle\int\limits_{0}^{1}dz \big( 4\alpha^{2}z^{4}(1-z^{3}) + 3 z^{3} (1-\alpha z^{2})^{2}\big)}{\displaystyle\int\limits_{0}^{1}dz \dfrac{z^{2}(1-z)}{(1+z+z^{2})}\bigg(1-\dfrac{b(\lambda^{2}|_{b=0})}{5}\zeta(z)\bigg)(1-\alpha z^{2})^{2}}~. 
\label{eigenvalue_eq_m=0}
\end{eqnarray}
Thus we obtain
\begin{eqnarray}
\lambda_{\alpha}^{2} = \dfrac{60\bigg(\alpha - \dfrac{3}{4} - \dfrac{27\alpha^{2}}{40}\bigg)}{\bigg[(30\ln3 - 10\sqrt{3}\pi + 21)\alpha^{2} + (120\ln3 -130)\alpha + (30\ln3 + 10\sqrt{3}\pi - 90) + \dfrac{b (\lambda^{2}|_{b=0})}{5}} ~.~~~~~~\label{eigenvalue_solution_m2=0}\\ 
\bigg((30\ln3 + 10\sqrt{3}\pi - 85.91)\alpha^{2} + (-60\ln3 + 20\sqrt{3}\pi -48.14)\alpha + (72 - 60\ln3 + )\bigg)\bigg]~~~~~~~~  \nonumber
\end{eqnarray}
For $b=0$, the eigenvalue expression (\ref{eigenvalue_solution_m2=0}) reduces to the following form
\begin{eqnarray}
\lambda_{\alpha}^{2}|_{b=0} = \dfrac{60\bigg(\alpha - \dfrac{3}{4} - \dfrac{27\alpha^{2}}{40}\bigg)}{(30\ln3 - 10\sqrt{3}\pi + 21)\alpha^{2} + (120\ln3 -130)\alpha + (30\ln3 + 10\sqrt{3}\pi - 90)} 
\label{eigenvalue_solution_m=0_b=0}
\end{eqnarray}
which attains minima at $\alpha \approx 0.50775$. The minimum value of $\lambda_{\alpha}^{2}|_{b=0}$ is found to be
\begin{eqnarray}
\lambda_{\alpha_{min.}}^{2}|_{b=0} \approx 13.7674~.
\label{lambda_m=0_b=0}
\end{eqnarray} 
The critical temperature is then determined using eq.(\ref{Critical Temp Min.}) and reads
\begin{eqnarray}
T_{c} = \dfrac{3}{4\pi}\sqrt{\dfrac{\tilde{\rho}}{\lambda_{\alpha_{min.}}|_{b=0}}} \approx 0.1239\sqrt{\tilde{\rho}}~.
\label{critic_temp_m=0_b=0}
\end{eqnarray}
It is interesting to note that the critical temperature obtained in this case, with the BI parameter $b = 0$, is matching exactly with the critical temperature obtained for the holographic $p$-wave superconductor constructed out of the Einstein-Yang-Mills theory \cite{sgdr1}.\\
Note that in eq.(\ref{eigenvalue_eq_m=0}) we would now use $\lambda_{\alpha_{min.}}^{2}|_{b=0}$ in place of $\lambda^{2}|_{b=0}$ for successive computations of the eigenvalues for different values of the BI parameter $b$. In that case, we can write eq.(\ref{eigenvalue_solution_m2=0}) as below
 \begin{eqnarray}
\lambda_{\alpha}^{2} = \dfrac{60\bigg(\alpha - \dfrac{3}{4} - \dfrac{27\alpha^{2}}{40}\bigg)}{\bigg[(30\ln3 - 10\sqrt{3}\pi + 21)\alpha^{2} + (120\ln3 -130)\alpha + (30\ln3 + 10\sqrt{3}\pi - 90) + \dfrac{b (13.7674)}{5}} ~~~~~~~\label{eigenvalue_solution_m=0_withlb0}\\ 
\bigg((30\ln3 + 10\sqrt{3}\pi - 85.91)\alpha^{2} + (-60\ln3 + 20\sqrt{3}\pi -48.14)\alpha + (72 - 60\ln3 + )\bigg)\bigg]~~~~~  \nonumber
\end{eqnarray}
where we have substituted $\lambda^{2}|_{b=0} = 13.7674$~.\\
We now take some small value for BI parameter $b$ in eq.(\ref{eigenvalue_solution_m=0_withlb0}) and minimize it with respect to $\alpha$ to find the corresponding eigenvalue $\lambda^{2}_{\alpha_{min.}}|_{b \ne 0}$. We then determine the critical temperature using eq.(\ref{Critical Temp Min.}). The critical temperature $T_{c}$ for some values of the BI parameter $b$ are given in Table I.
\subsubsection*{\textbf{Case(II): $m^{2} = -3/16,~ \Delta = 3/4$}}
\noindent In this case as well, we shall first find out the critical temperature when there is no BI correction, that is, $b = 0$ and shall then provide the critical temperature for some small values of the BI parameter $b$. To do so, we first write the functions $p(z)$, $q(z)$ and $r(z)$ deduced from eq.(\ref{general functions}) for this case. These functions have the following form for the present case
\begin{eqnarray}
 p(z) = z^{3/2}(1-z^{3}) \hspace*{38mm}~ \nonumber \\ 
 q(z) = \dfrac{33}{16} z^{5/2} \hspace*{47mm}~ \label{m=-3/16 functions}\\
 r(z) = \dfrac{z^{3/2}(1-z)}{(1+z+z^{2})}\bigg(1-\dfrac{b(\lambda^{2}|_{b=0})}{5}\zeta(z)\bigg). \nonumber  
\end{eqnarray}
Now we use the same trial function $F_{\alpha}(z)$, as in the previous case, along with the above functions to find eigenvalue given by
 \begin{eqnarray}
\lambda_{\alpha}^{2} = \dfrac{\displaystyle\int\limits_{0}^{1}dz \big( 4\alpha^{2}z^{7/2}(1-z^{3}) + \dfrac{33}{16} z^{5/2} (1-\alpha z^{2})^{2}\big)}{\displaystyle\int\limits_{0}^{1}dz \dfrac{z^{3/2}(1-z)}{(1+z+z^{2})}\bigg(1-\dfrac{b(\lambda^{2}|_{b=0})}{5}\zeta(z)\bigg)(1-\alpha z^{2})^{2}}~. 
\label{eigenvalue_eq_m=-3/16}
\end{eqnarray}
Upon solving for the integrals in the above expression, we get 
\begin{eqnarray}
\lambda_{\alpha}^{2} = \dfrac{\dfrac{3465}{5040}\bigg(3780\alpha - 2970 - 3178\alpha^{2}\bigg)}{\mathcal{D}} ~. 
\label{eigenvalue_solution_m=-3/16}
\end{eqnarray}
where \\
\hspace*{-12mm}$\mathcal{D} =  \bigg[(-3465\ln3 + 3776)\alpha^{2} + (-3465\ln3 + 3465\sqrt{3}\pi -14916)\alpha + (1732.5\ln3 + 1732.5\sqrt{3}\pi - 1150)+\dfrac{b (\lambda^{2}|_{b=0})}{5}  \bigg((1732.5\ln3 + 1732.5\sqrt{3}\pi - 11234.1667)\alpha^{2} + (6930\ln3 - 7974.1538)\alpha + (1732.5\ln3 - 1732.5\sqrt{3}\pi + 7994)\bigg)\bigg]$.\\
To find out the critical temperature in this case, we again put in some small values for the BI parameter $b$ in the above expression for the eigenvalue and then we go on to minimize it with respect to $\alpha$. After finding corresponding minimum values $\lambda^{2}_{\alpha_{min.}}$, we use eq.(\ref{Critical Temp Min.}) to determine the critical temperature $T_{c}$.\\
\noindent In Table I, we have provided tabular summary for the critical temperatue with the Born-Infeld correction for both the cases we have discussed above. It should be noted that the presence of the BI parameter is weakening the critical temperature for both the cases.   
 
\begin{table}[h]
\begin{center}
\begin{tabular}{ |c| c| c|}
\hline
~~~~~~Born-Infeld parameter, $b$~~~~~~ & \multicolumn{2}{c|}{~~~The critical temperature, $T_{c}$ ~~~~~~}         \\
\hline
& ~~~~~~$m^{2}=0,~\Delta = 1$~~~~~~ &~~~~~ $m^{2}=-3/16,~\Delta = 3/4$ ~~~\\
\hline 
0.0 & 0.1239$\sqrt{\tilde{\rho}}$ & 0.1425$\sqrt{\tilde{\rho}}$\\
\hline
0.01 & 0.1221$\sqrt{\tilde{\rho}}$ & 0.1414$\sqrt{\tilde{\rho}}$\\
\hline 
0.02 &  0.1201$\sqrt{\tilde{\rho}}$ & 0.1402$\sqrt{\tilde{\rho}}$ \\
\hline
0.03 &  0.1182$\sqrt{\tilde{\rho}}$ & 0.1390$\sqrt{\tilde{\rho}}$\\
\hline 
\end{tabular}
\label{tab1}
\end{center}
\caption{Critical temperature with the Born-Infeld correction}
\end{table}

\subsection{Condensation Operator}
\noindent Now we move on to find the condensation operator value. To calculate it we notice that near the critical temperature, we have $\rho (z)$ given by eq.(\ref{rho_sturm}). We have also found the solution for the field $\Phi(z)$ at the critical temperature $T_{c}$  (eq.(\ref{exact_Phi})). Now we expect that near the critical temperature, $\Phi(z)$ would slightly differ from eq.(\ref{exact_Phi}). For this reason, we add a small fluctuation $\chi(z)$ in the solution given in eq.(\ref{exact_Phi}) with appropriate boundary conditions. Hence, we have 
\begin{eqnarray}
\Phi(z) = \lambda r_{0}(1-z)\bigg[1-\dfrac{b(\lambda^{2}|_{b=0})}{10}\zeta(z)\bigg] + \dfrac{\langle \mathcal{O}_{\Delta} \rangle^{2}}{r_{0}^{2\Delta -1}}\chi(z)
\label{phi_strum}
\end{eqnarray}
where $\chi(1) = 0$ and $\chi^{\prime}(1) = 0$. \\
To determine the specific form of the field $\Phi(z)$ near the critical temperature, we substitute eq.(\ref{phi_strum}) in eq.(\ref{eom_Phi_in_z}) keeping terms only of $\mathcal{O}(b)$ and $\mathcal{O}(\langle \mathcal{O}_{\Delta} \rangle^{2})$. This gives the following equation for the fluctuation field $\chi(z)$
\begin{eqnarray}
\chi^{\prime\prime}+ 6b\lambda^{2}z^{3}\chi^{\prime} = \dfrac{\lambda z^{2\Delta} F^{2}}{r_{0}^{2}(1+z+z^2)}\bigg[1-\dfrac{b}{2}\bigg(\dfrac{(\lambda|_{b=0})^{2}}{5}\zeta(z)+3\lambda^{2}z^{4}\bigg)\bigg]~.
\label{eom_chi}
\end{eqnarray}
As the BI parameter $b$ is very small, we approximate $\lambda^{2}$ in eq.(\ref{eom_chi}) with $(\lambda|_{b=0})^{2}$ whenever it appears with $b$. In that case, eq.(\ref{eom_chi}) reduces to 
\begin{eqnarray}
\chi^{\prime\prime}+ 6b(\lambda|_{b=0})^{2}z^{3}\chi^{\prime} = \dfrac{\lambda z^{2\Delta} F^{2}}{r_{0}^{2}(1+z+z^2)}\bigg[1-\dfrac{b}{2}(\lambda|_{b=0})^{2} \bigg(\dfrac{\zeta(z)}{5}+3z^{4}\bigg)\bigg]~.
\label{eom_chi_1}
\end{eqnarray}
To find the solution of the above equation, we multiply it with $e^{\bigg(\dfrac{3b}{2}(\lambda|_{b=0})^{2}z^{4}\bigg)}$ and simplify it further to get the following form
\begin{eqnarray}
\bigg(e^{\bigg(\dfrac{3b}{2}(\lambda|_{b=0})^{2}z^{4}\bigg)}\chi^{\prime}\bigg)^{\prime} = \dfrac{\lambda z^{2\Delta} F^{2}}{r_{0}^{2}(1+z+z^2)}\bigg[1-\dfrac{b}{2}(\lambda|_{b=0})^{2} \bigg(\dfrac{\zeta(z)}{5}+3z^{4}\bigg)\bigg] e^{\bigg(\dfrac{3b}{2}(\lambda|_{b=0})^{2}z^{4}\bigg)}~.
\label{eom_chi_2}
\end{eqnarray}
Integrating eq.(\ref{eom_chi_2}) between $z=0$ and $z=1$ with the boundary conditions on $\chi(z)$ and $\chi^{\prime}(z)$, we find the following condition on the fluctuation field near the AdS boundary 
\begin{eqnarray}
\chi^{\prime}(0) = -\dfrac{\lambda}{r_{0}^{2}}\mathcal{A}_{\Delta}
\label{chi_prime_near_0}
\end{eqnarray}
where 
\begin{eqnarray}
\mathcal{A}_{\Delta} = \displaystyle\int\limits_{0}^{1}dz \dfrac{z^{2\Delta} F^{2}}{(1+z+z^2)}\bigg[1-\dfrac{b}{2}(\lambda|_{b=0})^{2} \bigg(\dfrac{\zeta(z)}{5}+3z^{4}\bigg)\bigg] \exp\bigg(\dfrac{3b}{2}(\lambda|_{b=0})^{2}z^{4}\bigg)~.
\label{Sturm Coeffi}
\end{eqnarray}
Taylor expanding $\chi(z)$ near the AdS boundary
\begin{eqnarray}
\chi(z) = \chi(0)+z\chi^{\prime}(0)+...
\label{chi expansion}
\end{eqnarray}
and comparing the coefficients of $z$ of eq.(s)(\ref{phi_strum}, \ref{phi_asymptote}) considering the above expansion of the field $\chi(z)$, we get
\begin{eqnarray}
- \dfrac{\tilde{\rho}}{r_{0}} = -\lambda r_{0}+\dfrac{\langle \mathcal{O}_{\Delta} \rangle^{2}}{r_{0}^{2\Delta -1}}\chi^{\prime}(0)~.
\label{rho chi 0 rel}
\end{eqnarray} 
Now we use eq.(\ref{chi_prime_near_0}) to substitute for $\chi^{\prime}(0)$ in the above equation. This yields
\begin{eqnarray}
 \dfrac{\tilde{\rho}}{r_{0}^{2}} = \lambda \bigg(1+\dfrac{\langle \mathcal{O}_{\Delta} \rangle^{2}}{r_{0}^{2\Delta +2}}\mathcal{A}_{\Delta}\bigg)~.
\label{operator_value_r0}
\end{eqnarray}
Finally we replace $r_{0}$ in terms of the Hawking temperature $T$ using eq.(\ref{Hawking_temperature}) and $\lambda$ in terms of the critical temperature $T_{c}$ using the relation $\lambda = \dfrac{\tilde{\rho}}{r_{0c}^{2}}$. This gives the condensation operator  in the following form
\begin{eqnarray}
\dfrac{\langle \mathcal{O}_{\Delta} \rangle}{T_{c}^{(\Delta+1)}} = \sqrt{\dfrac{2}{\mathcal{A}_{\Delta}}}\bigg(\dfrac{4\pi}{3}\bigg)^{(\Delta+1)}\sqrt{\bigg(1-\dfrac{T}{T_{c}}\bigg)}~.
\label{condensation_op}
\end{eqnarray}
In the above result $\Delta$ can take any positive value consistent with eq.(s)(\ref{con_dim}, \ref{B_F_bound}). It is also important to note that the condensation operator shows the second order phase transition  with the critical exponent $1/2$.\\
\noindent We have discussed two particular cases by choosing $m^{2}$ and the corresponding value for the conformal dimension $\Delta$ in the previous section. For those cases, the expression for the value of the condensation operator is given below. 
\subsubsection{\textbf{$m^{2} = 0,~ \Delta = 1$}}
\noindent In this case, eq.(\ref{condensation_op}) reduces to the following form
\begin{eqnarray}
\dfrac{\langle \mathcal{O}_{1} \rangle}{T_{c}^{2}} = \sqrt{\dfrac{2}{\mathcal{A}_{1}}}\bigg(\dfrac{4\pi}{3}\bigg)^{2}\sqrt{\bigg(1-\dfrac{T}{T_{c}}\bigg)}
\label{condensation_op_m=0}
\end{eqnarray}
where 
\begin{eqnarray}
\mathcal{A}_{1} = \displaystyle\int\limits_{0}^{1}dz \dfrac{z^{2} F^{2}}{(1+z+z^2)}\bigg[1-\dfrac{b}{2}(\lambda|_{b=0})^{2} \bigg(\dfrac{\zeta(z)}{5}+3z^{4}\bigg)\bigg] \exp\bigg(\dfrac{3b}{2}(\lambda|_{b=0})^{2}z^{4}\bigg)~.
\label{Sturm Coeffi_m=0}
\end{eqnarray}
Now we find the value of $\dfrac{\langle \mathcal{O}_{1} \rangle}{T_{c}^{2}}$ near $T \rightarrow 0$ such that eq.(\ref{condensation_op_m=0}) gives
\begin{eqnarray}
\dfrac{\langle \mathcal{O}_{1} \rangle}{T_{c}^{2}} \simeq \sqrt{\dfrac{2}{\mathcal{A}_{1}}}\bigg(\dfrac{4\pi}{3}\bigg)^{2} \approx \dfrac{24.8137}{\sqrt{\mathcal{A}_{1}}}
\label{condensation_op_m=0}
\end{eqnarray}
Taking the trial function $F_{\alpha} = (1 - \alpha z^{2})$ with the value of $\alpha$ that minimizes the eigenvalue $\lambda^{2}_{\alpha_{min.}}$ in $\mathcal{A}_{1}$ given by eq.(\ref{Sturm Coeffi_m=0}), we get 
\begin{eqnarray}
\mathcal{A}_{1} = \displaystyle\int\limits_{0}^{1}dz \dfrac{z^{2} (1 - \alpha z^{2})^{2}}{(1+z+z^2)}\bigg[1-\dfrac{b}{2}(\lambda|_{b=0})^{2} \bigg(\dfrac{\zeta(z)}{5}+3z^{4}\bigg)\bigg] \exp\bigg(\dfrac{3b}{2}(\lambda|_{b=0})^{2}z^{4}\bigg)~.
\label{Sturm Coeffi_m=0_trial}
\end{eqnarray}
We first consider the case when $b = 0$. In this case, eq.(\ref{Sturm Coeffi_m=0_trial}) becomes
\begin{eqnarray}
\mathcal{A}_{1} = \displaystyle\int\limits_{0}^{1}dz \dfrac{z^{2} (1 - 0.50775 z^{2})^{2}}{(1+z+z^2)}~.
\label{Sturm Coeffi_m=0_trial_b=0}
\end{eqnarray}
In the above eq.(\ref{Sturm Coeffi_m=0_trial_b=0}) we have used the value $\alpha \approx 0.50775$ which we have obtained in the previous section. We have shown there that at this value of $\alpha$, the eigenvalue attains its minimum value, $\lambda^{2}_{\alpha_{min.}}|_{b=0} \approx 13.7674$, when there is no BI correction. Using eq.(\ref{Sturm Coeffi_m=0_trial_b=0}) in eq.(\ref{condensation_op_m=0}), we find the value of $\dfrac{\langle \mathcal{O}_{1} \rangle}{T_{c}^{2}}$ is approximately 87.2482.\\     
We have also considered the BI correction to the value of the condensation operator. These corrections are listed in Table II for some small values of the BI parameter $b$.

\subsubsection{\textbf{$m^{2} = -3/16,~ \Delta = 3/4$}}
\noindent We now present the value of the condensation operator for $m^{2} = -\dfrac{3}{16}$ and $\Delta = \dfrac{3}{4}$. From eq.(\ref{condensation_op}), we get the following form of the condensation operator value in the present case
\begin{eqnarray}
\dfrac{\langle \mathcal{O}_{3/4} \rangle}{T_{c}^{7/4}} = \sqrt{\dfrac{2}{\mathcal{A}_{3/4}}}\bigg(\dfrac{4\pi}{3}\bigg)^{7/4}\sqrt{\bigg(1-\dfrac{T}{T_{c}}\bigg)}
\label{condensation_op_m=-3/16}
\end{eqnarray}
where 
\begin{eqnarray}
\mathcal{A}_{3/4} = \displaystyle\int\limits_{0}^{1}dz \dfrac{z^{3/2} F^{2}}{(1+z+z^2)}\bigg[1-\dfrac{b}{2}(\lambda|_{b=0})^{2} \bigg(\dfrac{\zeta(z)}{5}+3z^{4}\bigg)\bigg] \exp\bigg(\dfrac{3b}{2}(\lambda|_{b=0})^{2}z^{4}\bigg)~.
\label{Sturm Coeffi_m=-3/16}
\end{eqnarray}
As in the previous case, we find that near $T \rightarrow 0$ the value of $\dfrac{\langle \mathcal{O}_{3/4} \rangle}{T_{c}^{7/4}}$ is
\begin{eqnarray}
\dfrac{\langle \mathcal{O}_{3/4} \rangle}{T_{c}^{7/4}} \simeq \sqrt{\dfrac{2}{\mathcal{A}_{3/4}}}\bigg(\dfrac{4\pi}{3}\bigg)^{7/4} \approx \dfrac{17.3448}{\sqrt{\mathcal{A}_{3/4}}}~.
\label{condensation_op_m=-3/16}
\end{eqnarray}
We now take the trial function $F_{\alpha} = (1 - \alpha z^{2})$ with the value of $\alpha$ that minimizes the eigenvalue $\lambda^{2}_{\alpha_{min.}}$ in $\mathcal{A}_{3/4}$ given by eq.(\ref{Sturm Coeffi_m=0}) which gives 
\begin{eqnarray}
\mathcal{A}_{3/4} = \displaystyle\int\limits_{0}^{1}dz \dfrac{z^{3/2} (1 - \alpha z^{2})^{2}}{(1+z+z^2)}\bigg[1-\dfrac{b}{2}(\lambda|_{b=0})^{2} \bigg(\dfrac{\zeta(z)}{5}+3z^{4}\bigg)\bigg] \exp\bigg(\dfrac{3b}{2}(\lambda|_{b=0})^{2}z^{4}\bigg)~.
\label{Sturm Coeffi_m=-3/16_trial}
\end{eqnarray}    
We have considered the BI correction to the value of the condensation operator in this case as well which are listed in Table II for some small values of the BI parameter $b$.\\
\noindent In the table II, we display the value of condensation operator near $T = 0$ for two different cases ($m^{2}=0, \Delta = 1$) and ($m^{2}=-3/16, \Delta = 3/4$). We have noted earlier in table I that the critical temperature $T_{c}$ matches exactly for both the holographic $p$-wave superconductor models for the case ($m^{2}=0, \Delta = 1$) when the BI parameter $b$ is zero. However, the value of the condensation operator given in table II shows a departure by a factor of $\sqrt{2}$ from the value of condensation operator obtained in the Einstein-Yang-Mills $p$-wave holographic superconductor \cite{sgdr1}. It is also worth noting that the BI correction is increasing the values of the condensation operator in both the cases we have discussed.

\begin{table}[h]
\begin{center}
\begin{tabular}{ |c| c| c|}
\hline
~~~~~~Born-Infeld parameter ($b$)~~~~~~ & \multicolumn{2}{c|}{~~~The condensation operator value, $\langle \mathcal{O}_{\Delta} \rangle / T_{c}^{\Delta+1}$ ~~~~~~}         \\
\hline
& ~~~~~~$m^{2}=0,~\Delta = 1$~~~~~~ &~~~~~ $m^{2}=-3/16,~\Delta = 3/4$ ~~~\\
\hline 
0.0 & 87.2482 & 49.509\\
\hline
0.01 & 89.4636 & 50.1235\\
\hline 
0.02 &  92.5642 & 50.8645 \\
\hline
0.03 &  96.9787 & 51.7611\\
\hline 
\end{tabular}
\label{tab2}
\end{center}
\caption{Condensation operator value for different values of BI parameter}
\end{table}
\section{Conductivity}
\noindent In this section, we obtain the holographic conductivity,  which is accomplished by perturbing the gauge field in the bulk along the boundary, as a function of frequency. We consider the perturbation in the gauge field along $y$-direction $$A_{\mu} = (0,0,\phi(r,t),0)$$ 
where $\phi(r,t) = A(r)~e^{-i\omega t}$. However, we take the previous ansatz for the matter field which is given by $$\rho_\mu=(0,\rho(r),0,0)$$
Varying the action $\mathcal{S}$ in eq.(\ref{Action}) with respect to $A(r)$ and ignoring terms of $\mathcal{O}(b^{2})$ and $\mathcal{O}(\omega^{2}b)$, we get the following equation of motion corresponding to $A(r)$
\begin{eqnarray}
\hspace*{-15mm}\bigg(1-\dfrac{3b}{2r^{2}}f(r)A^{\prime2}\bigg)A^{\prime\prime} + \dfrac{f^{\prime}(r)}{f(r)}\bigg(1-\dfrac{b}{r^{2}}f(r)A^{\prime2}\bigg)A^{\prime} \nonumber \\
+\dfrac{b}{r^{3}}f(r)A^{\prime3}+\bigg(\dfrac{\omega^{2}}{f^{2}(r)}-\dfrac{2\rho^{2}}{r^{2}f(r)}\bigg)A=0 \hspace*{15mm}~
\label{eom_A_r_b}
\end{eqnarray}
where prime denotes derivative with respect to $r$. 
 Eq.(\ref{eom_A_r_b}) is highly nonlinear and is very difficult to solve. So for simplicity, we would ignore all the nonlinear terms in eq.(\ref{eom_A_r_b}). This can be done because nonlinear terms in eq.(\ref{eom_A_r_b}) appear with the BI parameter $b$ which is very small. However, one should note that the effect of the BI parameter would still enter in the solution through $\rho(z)$ which we have found in the previous section.  We shall now solve eq.(\ref{eom_A_r_b}) after ignoring all the nonlinear terms. This gives
\begin{eqnarray}
A^{\prime\prime} + \dfrac{f^{\prime}(r)}{f(r)}A^{\prime}+\bigg(\dfrac{\omega^{2}}{f^{2}(r)}-\dfrac{2\rho^{2}}{r^{2}f(r)}\bigg)A=0~.
\label{eom_A_r_c}
\end{eqnarray}
Now changing the coordinate to $z=\dfrac{r_{0}}{r}$, eq.(\ref{eom_A_r_c}) becomes
\begin{eqnarray}
A^{\prime\prime} + \bigg(\dfrac{f^{\prime}(z)}{f(z)}+\dfrac{2}{z}\bigg) A^{\prime}+\dfrac{r_{0}^{2}}{z^{4}}\bigg(\dfrac{\omega^{2}}{f^{2}(z)}-\dfrac{2z^{2}\rho^{2}}{r_{0}^{2}f(z)}\bigg)A=0~.
\label{eom_A_z_c}
\end{eqnarray}
We now move to the Tortoise coordinate given by 
\begin{eqnarray}
r_{*} = \int \dfrac{dr}{f(r)}
\label{tortoise_r}
\end{eqnarray}
which can be written in the $z$ coordinate as
\begin{eqnarray}
r_{*} = -\int \dfrac{dz}{r_{0}(1-z^{3})}~.
 \label{tortoise_z}
\end{eqnarray}
From eq.(\ref{tortoise_z}) we find that
\begin{eqnarray}
r_{*} = -\dfrac{1}{r_{0}}\bigg(\ln(1+z+z^2)^{1/6} - \ln (1-z^3)^{1/3}\bigg) - \dfrac{1}{\sqrt{3}r_{0}}\arctan\bigg(\dfrac{1+2z}{\sqrt{3}}\bigg) ~.
\label{r_star_full}
\end{eqnarray}
In that above equation the integration constant is chosen so that the AdS boundary appears at $r_{*} = 0$.
Considering leading order behaviour of eq.(\ref{r_star_full}), we get
\begin{eqnarray}
r_{*} \simeq ln (1-z)^{1/3r_{0}}~.
\label{r_star}
\end{eqnarray}
In the Tortoise coordinate, eq.(\ref{eom_A_r_c}) leads to the following equation 
\begin{eqnarray}
\dfrac{d^{2}A}{dr_{*}^{2}}+(\omega^{2}-V)A=0
\label{eom_A_r_star}
\end{eqnarray}
where $V$ is given by
\begin{eqnarray}
V=2(1-z^{3})\rho^{2}~.
\label{V_rho}
\end{eqnarray}
The solution to the above equation for $V = 0$ is straightforward and is given by
\begin{eqnarray}
A \sim e^{-i\omega r_{*}}~.
\label{V=0 solution}
\end{eqnarray} 
Using eq.(\ref{r_star}) in the above solution, we get  
\begin{eqnarray}
A \sim (1-z)^{-i\omega/3r_{0}}~.
\label{A_r_star_V_0}
\end{eqnarray}
We shall now generalize this solution for the case $V \neq 0$. In this case, we obtain
\begin{eqnarray}
A(z) = (1-z)^{-(i\sqrt{\omega^{2} - \langle V \rangle})/3r_{0}}
\label{general_A_z}
\end{eqnarray} 
where $\langle V \rangle$ is defined as
\begin{eqnarray}
\langle V \rangle = \dfrac{\int dr_{*} V |A(r_{*})|^{2}}{\int dr_{*}  |A(r_{*})|^{2}}~.
\label{<V>}
\end{eqnarray}
Now using eq.(\ref{rho_sturm}), with $F(z) \simeq 1$ near the boundary, we get
\begin{eqnarray}
V = (1-z^{3})z^{2\Delta}\dfrac{\langle \mathcal{O}_{\Delta} \rangle^{2}}{r_{0}^{2\Delta}}~.
\label{V_z}
\end{eqnarray}
Now using $V$ from eq.(\ref{V_z}) in eq.(\ref{<V>}) and using the fact that $r_{*} = -\dfrac{z}{r_{0}}$ near the boundary, we get the following expression for $\langle V \rangle$
\begin{eqnarray}
\langle V \rangle =\dfrac{\langle \mathcal{O}_{\Delta} \rangle^{2}}{2^{2\Delta}} \bigg(\dfrac{\Gamma (2\Delta +1)}{(-i\sqrt{\omega^{2} - \langle V \rangle})^{2\Delta}}\bigg)~.
\label{<V>2}
\end{eqnarray}
At low frequency, we can set $\omega = 0$, which leads to
\begin{eqnarray}
\langle V \rangle^{\Delta +1} = \dfrac{\langle \mathcal{O}_{\Delta} \rangle^{2}}{2^{2\Delta}} \Gamma (2\Delta +1)~. 
\label{<V>final}
\end{eqnarray}
For the two choices of $\Delta$ that we made earlier, we have the following expressions for $\langle V \rangle$
\begin{eqnarray}
\Delta = 1 \hspace*{10mm} \longrightarrow \hspace*{10mm} \langle V \rangle = \dfrac{\langle \mathcal{O}_{1} \rangle}{\sqrt{2}} \hspace*{19mm}~
\label{V_for_m=0}
\end{eqnarray}
\begin{eqnarray}
\Delta = \dfrac{3}{4} \hspace*{10mm}  \longrightarrow \hspace*{10mm} \langle V \rangle = \dfrac{3\langle \mathcal{O}_{3/4} \rangle^{8/7}}{8}\sqrt{\dfrac{\pi}{2}}~.
\label{V_for_m=-3/16}
\end{eqnarray}
Near $z \rightarrow 0$, we can expand $A(z)$ in eq.(\ref{general_A_z}) as  
\begin{eqnarray}
A(z) \simeq A(0)+zA^{\prime}(0)+\mathcal{O}(z^{2})+...~.
\label{A_z_near_0_a}
\end{eqnarray}
On the other hand, we know that we can expand gauge field near $z \rightarrow 0$ in the following manner
\begin{eqnarray}
A_{x}(z) \simeq A_{x}^{(0)}+\dfrac{A_{x}^{(1)}}{r_{0}}z +...~.
\label{A_z_near_0_b}
\end{eqnarray}
Now comparing eq.(s)(\ref{A_z_near_0_a}, \ref{A_z_near_0_b}), we get the following relations
\begin{eqnarray}
A_{x}^{(0)} = A(0)~~,~~ A_{x}^{(1)} = r_{0} A^{\prime}(0)~.
\label{related A's}
\end{eqnarray}
The expression for conductivity reads $$\sigma(\omega) = \dfrac{\langle J_{x} \rangle}{E_{x}} = -\dfrac{iA_{x}^{(1)}}{\omega A_{x}^{(0)}}~.$$
Then using eq.(\ref{related A's}), we get the following expression for conductivity  
\begin{eqnarray}
\sigma(\omega) = -\dfrac{ir_{0}}{\omega} \dfrac{ A^{\prime}(0)}{A(0)}~. 
\label{sigma_def}
\end{eqnarray}
Using eq.(\ref{general_A_z}) in eq.(\ref{sigma_def}), we find that 
\begin{eqnarray}
\sigma(\omega) = \dfrac{1}{3}\sqrt{1 - \dfrac{\langle V \rangle}{\omega^{2}}}~.
\label{general_sigma}
\end{eqnarray}
Substituting $\langle V \rangle$ from eq.(\ref{<V>final}) in eq.(\ref{general_sigma}), we obtain the following expression for the conductivity in the low frequency limit,
\begin{eqnarray}
\sigma(\omega) = \dfrac{i}{3} \dfrac{\langle \mathcal{O}_{\Delta} \rangle^{1/(\Delta+1)}}{2^{\Delta/(\Delta+1)}}\dfrac{\Gamma (2\Delta+1)^{1/2(\Delta+1)}}{\omega}~.
\label{Low_Freq_Conductivity}
\end{eqnarray}  
It is clear from eq.(\ref{Low_Freq_Conductivity}) that $\sigma(\omega)$ has a pole of order one. This implies that the DC conductivity diverges in this holographic $p$-wave superconductor model. Explicit expressions for DC conductivity for the cases we have considered in this paper are the following
\begin{eqnarray}
\Delta = 1 \hspace*{10mm}  \longrightarrow \hspace*{10mm} \sigma(\omega) = \dfrac{i}{3\omega}\sqrt{\dfrac{\langle \mathcal{O}_{1} \rangle}{\sqrt{2}}} \hspace*{19mm}~
\label{sigma_m=0}
\end{eqnarray} 
\begin{eqnarray}
\Delta = \dfrac{3}{4} \hspace*{10mm}  \longrightarrow \hspace*{10mm} \sigma(\omega) = \dfrac{i}{3\omega} \bigg(\dfrac{9\pi\langle \mathcal{O}_{3/4} \rangle^4}{128}\bigg)^{1/7}~.
\label{sigma_m=-3/16}
\end{eqnarray}

\section{Conclusions}
 
\noindent In this paper, we have studied a holographic model of a $p$-wave superconductor constructed from a massive vector field with the  nonlinear Born-Infeld electrodynamics in the matter sector of the Lagrangian. Considering probe approximation, where matter does not backreact with the spacetime geometry of the background, we have worked with a planar Schwarzschild-AdS metric. We have observed that the condensation gets suppresed due to presence of the Born-Infeld parameter $b$. In fact, we have found that the critical temperature for two choices of $m^{2}$, that is, ($m^{2} = 0, ~ -3/16$) decreases, making condensation harder, as we increase the value of $b$. We have also analysed the effect of Born-Infeld parameter in the condensation operator value. It turns out that the Born-Infeld correction to the condensation operator value is very nontrivial. It is found that the value of the condensation operator increases with the increase in the value of $b$ for both choices of $m^{2}$. \\  
\noindent We would like to point out that for the choice of $m^{2} = 0$, our result for the critical temperature, without Born-Infeld correction,  matches with the earlier non-Abelian model of the holographic $p$-wave superconductor, which is conceptually very different with the model we have considered in this paper. However, as we have pointed out earlier that the value of the condensation operator is different from the earlier model of holographic $p$-wave superconductor \cite{sgdr1}. This is because the two theories are quite different in form at the level of the action, although both exhibit a $p$-wave characteristic. With these observations, we conclude that the presence of Born-Infeld parameter is making the condensation difficult in the holographic  $p$-wave superconductor model considered in this paper.\\ 
\noindent  We have finally calculated the conductivity following a self-consistent approach developed in \cite{st} and have explicitly shown that the DC conductivity in this model indeed diverges. We would like to stress that such an analysis was absent in the context of $p$-wave holographic superconductors. \\

\noindent {\bf{Acknowledgements}}: DG would like to thank DST-INSPIRE, Govt. of India for financial support. SG acknowledges the Visiting Associateship at Inter-University Centre for Astronomy and Astrophysics, Pune.

\end{document}